\documentclass[aps,prl,twocolumn,groupedaddress,amsmath,amssymb]{revtex4-1}
\usepackage{graphicx}
\usepackage{color}

\graphicspath{{./}{./Figs/}}

\newcommand{\BraKet}[3]{\left\langle #1 \middle| #2 \middle| #3 \right\rangle}

\newcommand{\be}[1]{\begin{eqnarray}{\label{e#1}}} 
\newcommand{\beq}{\begin{eqnarray}}
\newcommand{\eeq}{\end{eqnarray}} 

\newcommand{\hide}[1]{}
\newcommand{\Eq}[1]{\textcolor{blue}{{Eq.}\!\!~(\ref{#1})}}

\definecolor{myred}{rgb}  {0.5,0.0,0.0}

\begin{document}
\title{Coherent Wave Propagation in Multi-Mode systems with Correlated Noise}
\author{Yaxin Li$^1$}
\author{Doron Cohen$^2$}
\author{Tsampikos Kottos$^1$}
\address{$^1$Wave Transport in Complex Systems Lab, Physics Department, Wesleyan University, Middletown CT-06459, USA}
\address{$^2$Department of Physics, Ben-Gurion University of the Negev, Beer-Sheva 84105, Israel}
\date{\today}
\begin{abstract}
Imperfections in multimode systems lead to mode-mixing and interferences between propagating modes. Such disorder is typically characterized 
by a finite correlation time (in quantum evolution) or correlation length (in paraxial evolution). We show that the long-scale dynamics of an initial 
excitation that spread in mode space can be tailored by the coherent dynamics on short-scale. In particular we unveil a universal crossover from 
exponential to power-law ballistic-like decay of the initial mode. Our results have applications to various wave physics frameworks, ranging from 
multimode fiber optics to quantum dots and quantum biology.
\end{abstract}


\maketitle

{\it Introduction.-- } 
The prevalence of wave coherent transport in multimode systems in the presence of noisy enviroments is a research theme, 
with relevance to a range of physics frameworks. For example, in the frameworks of quantum electronics, optics or matter waves the quest to 
develop methods that control coherence in many-particle systems at the quantum limit has inspired new quantum computation and information 
technologies that are emerging the last years \cite{S08,DM03,AO14,ARS01}. Recently, in the seemingly remote field of quantum biology \cite{ECRAMCBF07,PHFCHWBE10,SIFW10,L11,E11,FSC11,ZCBK17}, researchers have also provided experimental evidence of wavelike (coherent) 
energy transfer in ``warm, wet and noisy" enviroments. Prominent example is the establishment of the important role of coherence in optimizing 
photosynthesis. Such findings triggered a number of tantalizing questions like the possible role of coherent (quantum) physics in brain functions, 
etc. It is natural, therefore, to ask weather there are universal designed principles that enforce coherence dynamics in various wave transport 
settings where dynamical disorder (noise) cannot be ignored.

The same basic question emerges, yet, in classical wave transport in the framework of fiber optics \cite{K00}. Optical fibers have revolutionize 
many modern technologies ranging from medical imaging and information-transfer technologies to modern communications. Along these lines, 
multi-mode fibers (MMFs) \cite{RASC14,XABRRC16,HK13,XABRRC17} have recently been exploited as alternatives to single mode fibers-- 
the latter experiencing information capacity limitations, imposed by amplifier noise and fiber non-linearities. What makes MMFs attractive is the 
possibility to utilize the multiple modes as extra degrees of freedom in order to carry additional information -- thus increasing the information 
capacity of a single fiber. On the counter-side, MMF suffer from mode coupling due to external perturbations (index fluctuations and fiber bending 
and twisting) and from polarization scrambling effects due to fiber imperfections 
(core ellipticity and eccentricity, bending etc.). Both effects cause crosstalk and interference between propagating signals in different modes/
polarizations. To make things worst, the fiber imperfections vary with the propagation distance $z$ (aka quenched disorder).
It is, therefore, imperative to develop theories that take into consideration the role of disorder in the modal (and polarization) mixing 
and provide a quantitative description of light transport in MMFs.

\begin{figure}
\includegraphics[width=1\linewidth, angle=0]{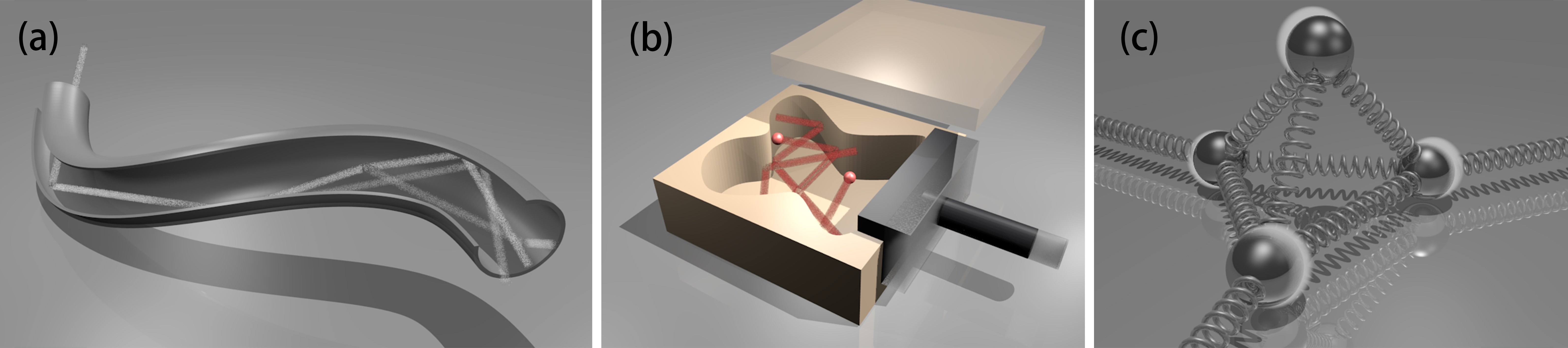}
\caption{(Color online) Schematics of various multi-mode systems in the presence of noisy environment: (a) A MMF experiencing twists, bendings,
and other forms of perturbations along the propagation direction $z$; (b) A multi-mode quantum dot (or a multi-mode opto-mechanical cavity) with 
an incoherently moving wall; (c) random network of coupled mechanical oscillators (slow envelope approximation) in the presence of noisy environment.
\label{fig1}}
\end{figure}

{\it Outline.-- }
In this paper we utilize a Random Matrix Theory (RMT) approach in order to unveil a physical mechanism that shields wave coherent effects in the 
presence of disorder. The RMT approach typically uncovers the most universal properties of wave transport in complex systems, and it can therefore 
serve as a good starting point for the understanding of designing schemes that protect the wave nature of propagation against noise. Specifically, 
we analyze the decay of an initial mode excitation (labeled $n_0$) in MMF that consist of $N$ modes with propagation constants $\beta_n=n\Delta$ 
where ${n=1,\cdots,N}$. The main objective is to study the decay of the survival probability $\mathcal{P}(z)$ towards its ergodic limit $\sim 1/N$. The 
mode mixing is due to quenched disorder associated with external perturbations along the propagation direction $z$ of the MMF. It is characterized 
by its strength $\varepsilon$ and by a correlation length $z_{c}$. From practical as well as physical point of view the interest is mainly in weak disorder ($\varepsilon<\sqrt{N}\Delta$), that can be characterized by a Fermi-Golden-Rule rate 
\beq \label{eGamma}
\Gamma \ = \ \frac{4\pi}{\Delta}\varepsilon^2 
\eeq
Consequently we distinguish between two length scales:
\beq
z_{\Delta}\equiv \frac{2\pi}{N\Delta};\quad\quad
z_{\Gamma} \equiv \frac{1}{\Gamma} 
\eeq 
The former is the short length scale over which the bandwidth is resolved, 
while the latter characterizes the non-stochastic coherent decay of an excitation.    
We distinguish between {\em short} correlation length ($z_c<z_{\Delta}$) and {\em long} correlation length ($z_c > z_{\Gamma}$). In the latter regime 
we find ballistic-like decay $\mathcal{P}(z) \sim 1/z$ as opposed to the exponential decay for shorter $z_c$. In the concluding paragraph we emphasize 
that the results of our study are relevant for a wide range of multi-mode or multi-level physical settings (see Fig. \ref{fig1}), appearing in areas as diverse 
as mesoscopic optics, and matter waves to quantum electronics and quantum biology.

{\it RMT modeling.-- } 
We presume that the perturbations along the propagation distance of the fiber induce only coupling between forward 
propagating modes (paraxial approximation). Furthermore, we consider that the fiber can be described in terms of concatenated 
segments of length $z_{c}$ which are associated with statistically independent fiber perturbations. Based on these assumptions we can write 
a paraxial Hamiltonian $H^{(k)} = H_{0} + \varepsilon B^{(k)}$  that describes the field propagation within the $k$-th segment.
Here $H_0$ describes the unperturbed fiber and $B^{(k)}=\left(B^{(k)}\right)^{\dagger}$ represents the perturbation of the $k$-th segment that 
is responsible for the mode mixing. In the mode representation the $N\times N$ matrix $H_0$ is diagonal with elements ${H_{nm}^{(0)}= \beta_n 
\delta_{n,m}}$. For simplicity we assume that the mode propagation constants are equally spaced, namely $\beta_n = n \Delta$ where ${n=1,
\cdots,N}$. The perturbation matrix $B^{(k)}$ is modeled as a GUE random matrix. For such matrix $\langle |B|^2\rangle=2$, hence the off 
diagonal terms of the Hamiltonian have dispersion $2\varepsilon^2$ and zero average. Note that this factor of 2 is reflected in the definition of 
\Eq{eGamma}.

The field propagation in each section $k$ is described by the unitary matrix
\begin{equation}
U^{(k)} \ = \ e^{-i\left(H_0+\varepsilon B^{(k)}\right) z_c} 
\label{Tmatrixk}
\end{equation}
In the analysis below, we do not consider polarization degrees of freedom. It can be shown that their presence does not alter the general 
picture (appart from an abrupt drop in the survival probability during the first evolution step), and therefore we omit them for a better clarity 
of the presentation. Below, unless stated otherwise, we assume that the paraxial distances are measured in units of mean propagation 
constant spacing $\Delta$. 

The one step dynamics is characterized by a stochastic kernel 
\beq
P(n|n_0) \ &=& \ \overline{\left|\BraKet{n}{U^{(k)}}{n_0}\right|^2} 
\\ \label{eL}
&\equiv& \ (1{-}\lambda) \delta_{n,n_0} + \lambda W(n-n_0) 
\eeq
Here we averaged the one-step dynamics over realizations of the random matrix $B^{(k)}$. The parameter $\lambda$ is defined as the 
probability that is drained from the initial mode after one step. The function $W(n-n_0)$ describes the distribution of the probability 
over the other modes.  
The modal field amplitudes $\Psi_n(z)$ at distance $z$ along the MMF are determined by operating 
on the initial state ${ \Psi_n(0) = \delta_{n,n_0} }$ with an ordered sequence of $U^{(k)}$ matrices (${k=1,2,\cdots}$).   
This multi-step dynamics generates a distribution ${ P_z(n|n_0) = |\Psi_n(z)|^2 }$. 
Below we discuss how    ${ P_z(n|n_0) }$ is related to ${ P(n|n_0) }$, 
and what are the implications regrading the survival probability 
\beq
\mathcal{P}(z) \ \ \equiv \ \ P_z(n_0|n_0) 
\eeq

\begin{figure}
\includegraphics[width=.85\linewidth, angle=0]{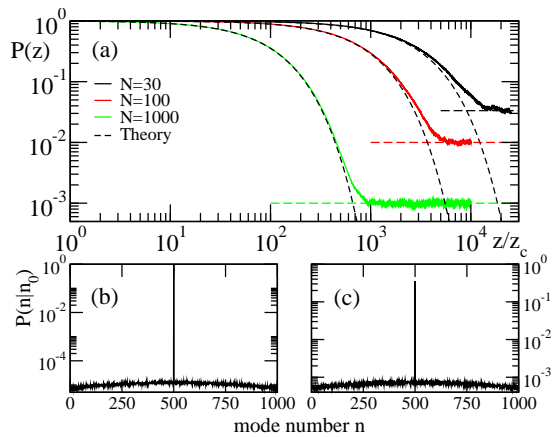}
\caption{(Color online) 
(a) The decay of the survival probability for short correlation lengths $z_c=0.005$, and perturbation strength $\varepsilon=0.5$. 
The units are chosen such that $\Delta=1$. The various colored curves indicate MMFs with different number of modes $N=30, 
100, 1000$.  The colored horizontal dashed lines indicate the ergodic value ${\mathcal{P}(z) \approx 1/N}$. The black dashed 
lines indicate Eq.(\ref{surv2}); (b)~The coherent spreading $P(n|n_0)$ for $z=z_c$; (c)~The spreading profile $P_z(n|n_0)$ for 
$z=100z_c$.  
\label{fig2}}
\end{figure}

{\it Short correlation length.-- }
For short segment ($z<z_{\Delta}$) the probability that is transferred to each of the $N$ modes is $2\varepsilon^2 z_c^2$ hence the total 
probability that is drained from the initial mode is 
\beq
\lambda \ = \ N \times 2\varepsilon^2 z_c^2 \ = \ \frac{z_c^2}{z_{\Gamma} z_{\Delta}}  
\eeq
As long as the first term in \Eq{eL} dominates, successive convolutions lead to exponential decay, namely, after $t$ steps ${ (1{-}\lambda)^t 
\approx \exp(-\lambda t) }$, with $t=z/z_c$, hence 
\begin{equation}
\label{surv2}
\mathcal{P}(z) \ = \ \exp\left[-\frac{\ z_c}{z_{\Delta} z_{\Gamma}} z \right] ,
\ \ \ \ \mbox{for $z_c<z_{\Delta}$}
\end{equation}
The above has been tested numerically and was found to reproduce nicely the results of our simulations for various $N$-values, see 
Fig. \ref{fig2}a. At the same figure we also display the single step $P(n|n_0)$, and the $P_z(n|n_0)$ distribution after 100 steps, see 
Fig. \ref{fig2}b and Fig. \ref{fig2}c respectively.
In both instances the shape of the evolving distribution is dominated by a delta peak around the initial mode ($n_0=N/2$).
This delta peak is gradually drained, until it attains the ergodic value ${\mathcal{P}(z) \approx 1/N}$.

\begin{figure}
\includegraphics[width=.85\linewidth, angle=0]{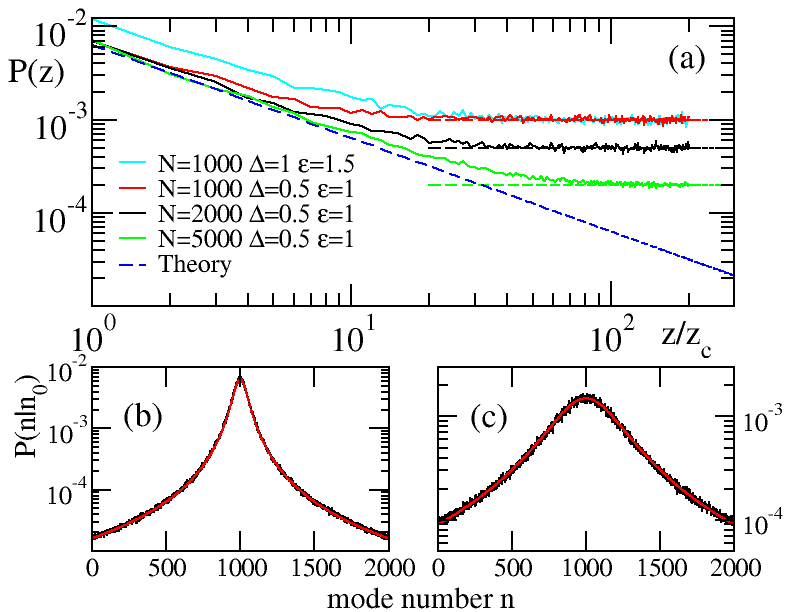}
\caption{(Color online) 
(a) The decay of the survival probability for large correlation length $z_c=0.32$. The various colored curves indicate MMFs with 
different parameters as indicated in the figure legend. The horizontal dashed lines indicate the ergodic limit ${\mathcal{P}(z)\approx
1/N}$.  The blue dashed line indicates Eq.(\ref{power}). (b)~The coherent spreading $P(n|n_0)$ for $z=z_c$; (c)~The spreading 
profile $P_z(n|n_0)$ for $z=5z_c$. In both (b,c) the parameters are $\Delta=0.5$, $\varepsilon=1$ and $N=2000$ while the red 
dashed line indicates the Lorentzian of Eq.(\ref{Lorentz}).
\label{fig3}}
\end{figure}

{\it Large correlation length.-- }
For $z_c > z_{\Gamma}$ it is well known from the study of the coherent dynamics \cite{CIK00,Mello97} that the 
initial delta peak completely dissolves, and one obtains \Eq{eL} with $\lambda\sim 1$  and Lorentzian line shape
\beq
\label{Lorentz}
W(n-n_0) = \frac{\Delta}{\pi} 
\frac{\Gamma} {[(n-n_0)\Delta]^2 + \Gamma^2}
\eeq
This line shape is obtained after distance $z_{\Gamma}$. After a larger distance  $z_c > z_{\Gamma}$ the line shape does not 
change, but the phases of the wavefunction are further randomized. It follows that the coherent evolution over successive segments 
can be approximated as a convolution of $W(n'-n'')$ kernels. We therefore get effectively stochastic evolution. But this stochastic 
evolution does not obey the central limit theorem. It is of the Levy-flight type because the Lorentzian does not have a finite second 
moment. Successive convolutions of $t=z/z_c$ Lorentzians give a wider Lorentzian of width $\Gamma t$. It follows from 
\Eq{Lorentz} that the survival provability decays in a ballistic-like fashion:
\begin{equation}
\label{power}
\mathcal{P}(z) \ = \  2  \frac{z_{\Gamma} z_c}{N z_{\Delta}} \frac{1}{z} ,  
\ \ \ \ \mbox{for $z_c > z_{\Gamma}$}
\end{equation}
The above picture is nicely confirmed by our detailed numerical analysis. In Fig. \ref{fig3}a we report our findings for the survival
probability for various mode sizes $N$, and perturbations strengths $\varepsilon$. In Fig. \ref{fig3}b we also report the Lorentzian 
waveform at the end of the coherent evolution $z=z_c$. The robustness of the Lorentzian shape Eq. (\ref{Lorentz}) against the 
dynamical disorder is further confirmed in Fig. \ref{fig3}c where we plot $P_z(n|n_0)$ after $5$ segments.

\begin{figure}
\includegraphics[width=.85\linewidth, angle=0]{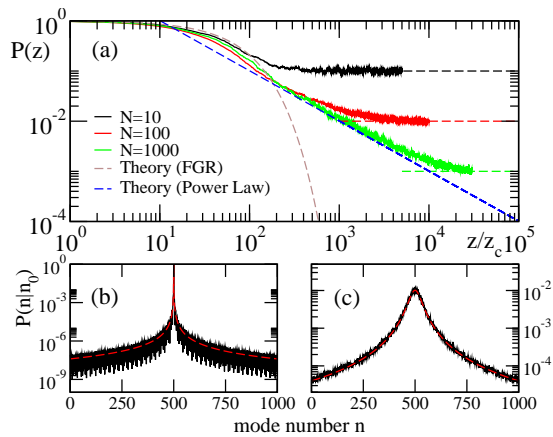}
\caption{(Color online) 
(a) The decay of the survival probability for intermediate correlation length $z_c=1$, and perturbation strength $\varepsilon=0.05$. 
The units are chosen such that $\Delta=1$. The various colored curves indicate MMFs with different number of modes ${N=10, 100, 
1000}$. The color horizontal dashed lines indicate the ergodic limit ${\mathcal{P}(z)\approx 1/N}$. The brown dashed line indicates 
the FGR exponential decay Eq.(\ref{FGR}), while the blue dashed line indicates the power-law decay Eq.(\ref{power}). (b)~The coherent 
spreading $P(n|n_0)$ for $z=z_c$; (c)~The spreading profile $P_z(n|n_0)$ for $z=1000z_c$. In both (b,c) the parameters are 
$\Delta=1$, $\varepsilon=0.05$ and $N=1000$ while the red 
dashed line indicates the Lorentzian of Eq.(\ref{Lorentz}).
\label{fig4}}
\end{figure}

{\it Intermediate correlation length.-- }
Consider $z_{\Delta}<z_c<z_{\Gamma}$. In this case, the initial spreading is dictated by a Fermi-Golden-Rule (FGR) type picture. 
Namely, the probability that is transferred to each of the modes within the unresolved bandwidth $2\pi/z_c$ is $(\varepsilon 
z_c)^2$, hence the total probability that is drained from the initial mode is 
\beq
\lambda \ \ = \ \ \Gamma z_c
\eeq
The analysis proceeds as in the discussion of short correlation scale, just with this different expression for $\lambda$. Namely, 
as long as the first term in \Eq{eL} dominates, successive convolutions lead to exponential decay ${ \exp(-\lambda t) }$ with 
$t=z/z_c$. Consequently we obtain a result that is independent of $z_c$, namely,  
\begin{equation}
\label{FGR}
\mathcal{P}(z) \ = \ \exp\left[- \frac{1}{z_{\Gamma}} z \right],
\ \ \ \ \mbox{for $z_{\Delta} < z_c < z_{\Gamma}$}
\end{equation}
Equation (\ref{FGR}) compares nicely with the numerical simulations, see Fig. \ref{fig4}a. Notice that as opposed to Eq. (\ref{surv2}), 
now the decay rate does not involve the number of modes of the system $N$ and neither depends on $z_c$. 
At the same time the envelope of the evolving waveform acquires Lorentzian-like tails spilled all over the $N$ modes, 
see Fig. \ref{fig4}b. Nevertheless, the dominant component of the waveform is centered at the initial mode $n_0$.
For larger propagation distances $z>z_{\Gamma}$, the FGR decay law Eq. (\ref{FGR}) cease to apply. Instead, either the waveform reach 
an ergodic distribution (see the black line in Fig. \ref{fig4}a, corresponding to $N=10$) or (in the case of large number of modes 
$N$) it continues spreading; albeit with a different form. Specifically, the previous argument associated with the robustness
of the Lorentzian waveform against noise takes over, and we recover the physics that led us to Eqs. (\ref{Lorentz},\ref{power}), see Fig.
\ref{fig4}a,c.

{\it Strong disorder, diffusive decay.-- }
So far we have discussed weak disorder. We now turn to discuss briefly the strong disorder regime ($\varepsilon > \sqrt{N}\Delta$).
The scenario for short ${z_c}$  is formally the same as that of the ``short correlation" analysis, leading to an exponential decay.
But if $z_c$ exceeds ${z_{N} = 1/(\sqrt{N} \varepsilon)}$ the probability is drained from the initial mode, and the distribution 
becomes ergodic with ${\mathcal{P}(z)\approx 1/N}$. At this stage one wonders why the naively expected diffusive decay does not 
appear. Are we missing something in the analysis? The answer is that the analysis so far has assumed $B$ that looks like a full 
GUE matrix. But in more general circumstance $B$ might have a finite bandwidth ${b \ll N}$. The analysis for the weak disorder 
regime still holds but with $N$ replaced by $b$. In contrast, in the strong disorder regime, it is well known \cite{CIK00} that the 
saturation profile is not a Lorentzian. Rather, if $z_c$ is long enough, the saturation profile is exponentially localized over 
${\xi = b^2}$ modes. Such saturation profile has a finite second moment. Consequently, the same argumentation as in the ``long 
correlation regime" implies that the width of the distribution evolves as $\xi \sqrt{t}$, where  $t=z/z_c$ is the number of steps. This 
leads to the conclusion that the survival provability decays in a diffusive-like fashion: 
\begin{equation}
\mathcal{P}(z) \ \approx \  \frac{1}{b^2} \sqrt{\frac{z_c}{z}}   
\end{equation}
Because of lack of space, we defer a more detail discussion of other results and a thorough analysis of the decay of the survival 
probability for the more realistic case where $b\ll N$ to a later publication \cite{LCK19}.

{\it Summary -- } 
We have illuminated the interplay between the short time coherent evolution and the long time stochastic spreading in multimode 
systems. The correlation scale $z_c$ of a disordered environment determines the crossover from an exponential-decay to diffusive-
like or ballistic-like decay. The latter is due to a Levy-type spreading which is implied by convolution of Lorentzian kernels.
Our results have been formulated using a universal RMT modeling. A future direction that we currently pursue \cite{LCK19} is to 
design other coupling schemes for which the one-step coherent evolution leads to tailored anomalous decay of the survival probability.

{\it Acknowledgments --} (Y.L) and (T.K) acknowledge partial support by an AFOSR grant No. FA 9550-10-1-0433, and by an NSF grant 
EFMA-1641109. Y.L. acknowledge funding from the Wesleyan University CIS summer research program. (D.C.) acknowledges support 
by the Israel Science Foundation (Grant No. 283/18). The authors acknowledge useful discussions with E. Makri and Y. Cai who participated 
at the initial stage of the project.


\end{document}